# Optical Activation of Germanium Plasmonic Antennas in the Mid Infrared

Marco P. Fischer[1], Christian Schmidt[1], Emilie Sakat[2], Johannes Stock[1], Antonio Samarelli[3], Jacopo Frigerio[4], Michele Ortolani[5], Douglas J. Paul[3], Giovanni Isella[4], Alfred Leitenstorfer[1], Paolo Biagioni[2], and Daniele Brida[1,*]

[1] Department of Physics and Center for Applied Photonics, University of Konstanz, D-78457 Konstanz, Germany

[2] Dipartimento di Fisica, Politecnico di Milano, Piazza Leonardo da Vinci 32, 20133 Milano, Italy

[3] School of Engineering, University of Glasgow, Rankine Building, Oakfield Avenue, Glasgow, G12 8LT, UK

[4] L-NESS, Dipartimento di Fisica del Politecnico di Milano, Via Anzani 42, 22100 Como, Italy

[5] Department of Physics, Sapienza University of Rome, Rome, 00185 Italy

[*] e-mail: daniele.brida@uni-konstanz.de

**Abstract** Impulsive interband excitation with femtosecond near-infrared pulses establishes a plasma response in intrinsic germanium structures fabricated on a silicon substrate. This direct approach activates the plasmonic resonance of the Ge structures and enables their use as optical antennas up to the mid-infrared spectral range. The optical switching lasts for hundreds of picoseconds until charge recombination red-shifts the plasma frequency. The full behavior of the structures is modeled by the electrodynamic response established by an electron-hole plasma in a regular array of antennas.

Plasmonics offers an elegant way to effectively couple optical radiation to sub-wavelength structures [1, 2, 3, 4, 5, 6]. Owing to the strong field enhancement, it becomes possible to access light-matter interactions at the nanometer scale [7, 8, 9, 10] with the opportunity to efficiently excite single quantum systems [11, 12]. In particular, the mid-infrared (MIR) spectral range is of interest for sensing explosives, hazardous chemicals and molecules of biological relevance in the so-called "vibrational fingerprint region" that covers the wavelength band from 3 to 20 μm [13, 14, 15, 16, 17], i.e. from 100 THz to 15 THz. In this context, Ge represents a novel material for MIR plasmonics [18, 19, 20]. Recent technological advancements enable ultra-high doping of single-crystalline films and the approach of growing Ge on Si substrates ensures full compatibility with standard semiconductor technologies. The doping level can be tailored in order to tune the carrier concentration and thus to control the plasma frequency $\nu_p$ [21, 22, 23]. State-of–the-art doping techniques allows semiconducting materials to be employed for plasmonics applications up to MIR frequencies [19].

Semiconductors are also appealing for the possibility to excite optically electrons from the valence to the conduction band, thus establishing a plasma response [24, 25] that lasts until the charges recombine. This approach has been successfully exploited to study active plasmonics [26, 27, 28] and THz metamaterial devices [29, 30] in direct bandgap semiconductors such as InSb, GaAs [31, 32, 33], or in Si [34]. The plasma frequencies were achieved in these materials, however, were limited to the far-infrared or terahertz ranges.

In this work, we demonstrate the activation of MIR plasmonic resonances in Ge microstructures by impulsively establishing a plasma response that extends up to a frequency of 60 THz with ultrashort near-infrared pulses. We also provide a theoretical insight on the plasmonic behavior of the antennas by modeling their optical response triggered by an electron-hole plasma in the semiconductor band structure.

Ge is the ideal material for our application since: i) it can be effectively excited by near-infrared radiation resonant with direct transitions; ii) the indirect gap prevents quick recombination of photoexcited carriers with the consequential drift of the plasma frequency on ultrafast timescales; iii) Ge has no dipole-active optical phonons in the MIR spectral region that complicate its dielectric behavior and speed up the recombination process [18, 24, 35].

When an ultrashort near-infrared pulse excites intrinsic Ge, electron-hole pairs are created in the Γ-valley of the band structure via interband absorption [see Fig. 1(a)]. Electrons at the Γ-point scatter into the L minima within 100 fs, where they acquire a larger effective mass $m^*$. For high pump fluence, two-photon absorption has also to be taken into account and allows access to the large joint density of states around the L symmetry point. The occurrence of photoexcited electrons in the L-valley of the conduction band and holes in the valence band at the Γ point leads to a Drude-like dielectric response with the intrinsic Ge becoming quasi-metallic. The resonance characteristics of optical structures depend significantly on both their geometry as well as the inherent number of free carriers. Thus, impulsive excitation can be used to activate and tune the resonance properties of Ge plasmonic antennas.

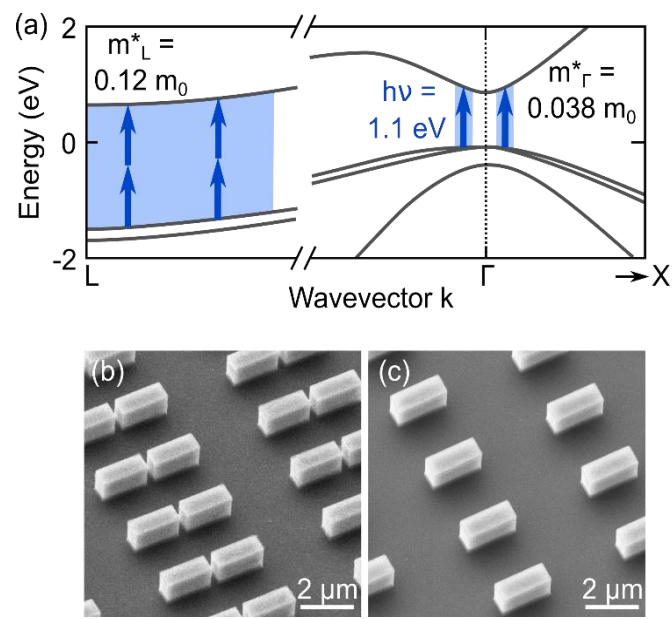

Figure 1 (color online): (a) Sketch of the band structure of Ge [35] with near-infrared transitions at the direct gap and two-photon absorption at the L symmetry point. The graph includes the effective masses for electrons in different valleys with respect to their rest mass $m_0$. Scanning electron micrographs of (b) a 2 µm double rod antenna array and (c) a 2 µm single rod antenna array fabricated from a 1µm thick, intrinsic Ge epitaxial layer grown on a Si substrate.

Single-crystalline, undoped Ge films were grown on intrinsic Si substrates by low-energy plasma-enhanced chemical vapor deposition [20, 23, 36, 37]. The 1-µm-thick layer was patterned by electron beam lithography using a Vistec VB6 tool with subsequent reactive ion dry etching [see Fig. 1(b) and (c)] using hydrogen silsesquioxane resist and a mixed $SF_6$ and $C_4F_8$ process [19, 38, 39]. With these steps we produced single and double rod antennas with a width of 800 nm and an arm length of either 2 µm or 3 µm. The double rod antennas consist of two equal arms featuring a gap size of 300 nm. The design of the arrays is optimized to provide minimum coupling between the localized plasmonic modes of different adjacent antennas while preserving the maximum coverage of the substrate.

The optical control of the Ge optical antennas is driven by an ultrafast laser system pumped by a Ti:sapphire regenerative amplifier [24]. It is seeded by a femtosecond Er:fiber laser that supplies also the 8 fs pulses used for electro-optical sampling (EOS) of the MIR transients. A two-stage non-collinear optical parametric amplifier (NOPA) [40] generates sub-20 fs pulses with energies up to 15 µJ at a central wavelength tuned to 1050 nm for the efficient excitation of direct interband transitions in the Ge structures. The large photon fluence also allows for two-photon absorption and the creation of a high concentration of free carriers that support a plasmonic behavior up to a frequency of approximately 60 THz [24]. Si absorbs the near-infrared wavelength only via inefficient indirect transitions resulting in a negligible photo-carrier density in the substrate [35].

To probe the MIR activation of the Ge antennas we employed broadband phase-stable pulses extending from 15 THz to 30 THz. These transients are generated via difference frequency generation between the output beams of two additional OPAs operating at center wavelengths of 1.18 µm and 1.28 µm, respectively. The nonlinear mixing takes place in a 250-µm-thick GaSe crystal to guarantee sufficient probing bandwidth and a pulse duration of 125 fs [41].

To trigger the switching dynamics of the Ge structures we focused collinearly the near-infrared pump and MIR probe pulses with a large-aperture parabolic mirror. The incidence on the sample was set at the Brewster's angle condition for the Si substrate, $\alpha_{Br} = 74°$. The probe electric field was selected to be *p*-polarized and parallel to the long antenna axis. The time interval $\Delta t$ between excitation and probing was adjusted with an optical delay stage. Between 2000 and 3000 antennas are excited at a focus diameter of 130 µm. The dimension of our structures is significantly larger than the pump wavelength of 1050 nm. Therefore, the near-infrared pump is absorbed in the Ge antennas following the conventional Lambert Beer's law [42] without significant geometry-related photonic effects. Instead, the MIR interaction with the antennas is strongly governed by the sub-wavelength size of the elements and the dynamical activation of their plasmonic resonance. The pump pulse intensity is controlled to optimally tune the maximum plasma frequency obtained in Ge without reaching the threshold for optical damage.

An off-axis parabolic mirror collimates the radiation reflected in the specular direction. This geometry is chosen to minimize any influence of the substrate. The MIR probe transients are then characterized in amplitude and phase by electro-optical sampling in a 90- µm-thick GaSe crystal [41]. The reflection spectra are obtained by Fourier transform. The excitation is modulated at 500 Hz, i.e. at half the repetition rate of the system. The probe is then measured in the presence and absence of the pump pulse in order to extract the transient reflectivity of the sample.

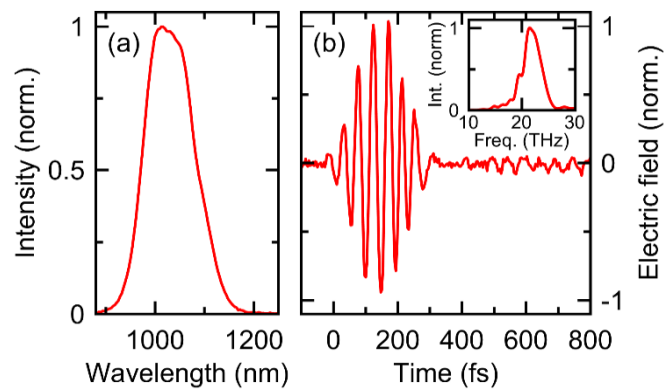

Figure 2 (color online): (a) The intensity spectrum of the near-infrared pulse that activates the Ge antennas. (b) The MIR probe transient as measured by electro-optical sampling and corresponding intensity spectrum (inset).

Without excitation, the antennas do not display any plasmonic behavior and we detect the MIR light reflected by the array. After near-infrared excitation, a significant reduction of the intensity in the specular direction is observed. Fig. 3(a) demonstrates the electric field profile of the MIR pulse as reflected by the sample before (black line) and after photoexcitation (red line). The spectral intensity is thus reduced by 65% as depicted in Fig. 3(b). This effect is due to the increased scattering and absorption cross section of the Ge antennas once the plasmonic resonance has been established. In fact, the probe light is mainly re-emitted in the direction normal to the sample surface rather at large angles as expected by a dipole radiation pattern.

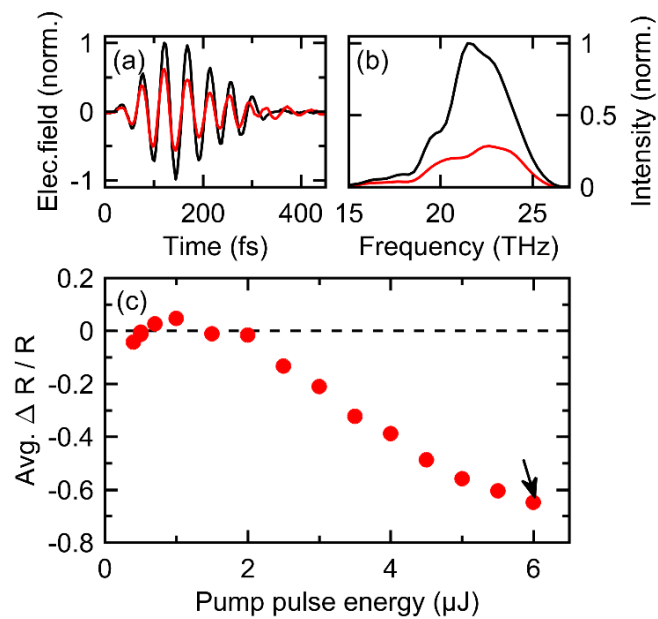

Figure 3 (color online): (a) The electric field time trace and (b) the spectral intensity of the MIR probe pulse reflected from the 2-μm double-rod antenna array with (red) and without (black) near-infrared optical excitation at a pump energy of 6 μJ. The carrier density in the active antenna is estimated to be approximately $1.6 \cdot 10^{20}$ cm$^{-3}$. (c) The integrated reflectivity change as a function of the pulse energy recorded at a delay time of 125 ps after excitation. The arrow indicates the experimental conditions corresponding to the data depicted in (a) and (b).

It should be noted that any residual pump-probe signal from the substrate would increase the reflectivity rather than diminishing it, as expected for the photoexcitation of an unstructured semiconductor material [24]. The pump-probe signal vanishes almost completely by rotating the samples by 90° to set the probe beam polarization perpendicular to the long antenna axes. This aspect further proves the activation of localized plasmons that become resonant within the geometrical constraints of the structures. Qualitatively, the switching action is similar for all the array samples investigated. The response we measured is particularly broadband as expected for plasmonic resonances established in proximity of the plasma frequency.

In Fig. 3(c) we plot the fluence dependence for the reflectivity of the 2 µm Ge double antenna array at a fixed delay between pump and probe of 125 ps. We notice that the reflectivity initially displays a slight increase. This is mainly due to the onset of a low plasma frequency in Ge. For higher pumping fluence, the antennas are plasmonically activated and interact with MIR photons. We can define a threshold for the switching at the fluence of 3 mJ/cm$^2$ (2 µJ pulse energy) that we estimate to create a plasma frequency of 20 THz [24].

Fig. 4(a) reports the transient reflectivity of the antennas array as a function of the time delay between near-infrared excitation and MIR probing. In a previous work [19], we studied the steady-state localized resonances in heavily-doped Ge antennas with the same geometry as the one investigated here. We demonstrated the occurrence of two distinct features, one with the near fields concentrated at the Ge-Si interface (lower-energy resonance) and one with the near fields concentrated at the Ge-air interface (higher-energy resonance). Here we can follow the evolution of the Ge-Si resonance after optical activation of the intrinsic antennas while the plasma frequency red-shifts as the carriers recombine in Ge. The relatively short recombination time, measured to be approximately 300 ps, is in good agreement with studies on thin films and nanowires where surface recombination plays a significant role [37, 43]. Directly after excitation, the induced plasma frequency and reaches a frequency of 60 THz. In these conditions, the Ge-air resonance lies at approximately 45 THz and therefore is outside our observation window. On the other side, we clearly observe the signature of the lower-energy resonance between 24 and 20 THz. Subsequently, the resonance redshifts to below 19 THz in about 400 ps, until no resonance characteristics are detectable anymore once $\nu_p$ becomes lower than 15 THz. This drift of the plasmonic response is due to electron-hole recombination and the consequent reduction of the carrier density.

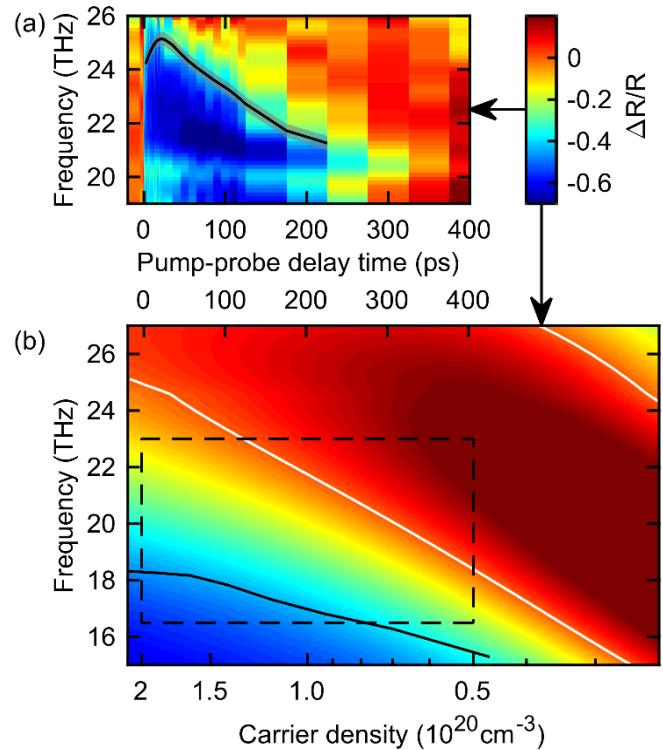

Figure 4 (color online): (a) The spectrally resolved relative reflectivity change (ΔR/R, color encoded) of the 2 µm single-rod antenna array under near-infrared excitation. The data was acquired at a 5 µJ pump pulse energy as a function of the delay time between the pump and the MIR probe. The black solid line follows the calculated Ge-Si resonance frequency and the shaded area represents the confidence interval of its value. (b) Simulation of the reflectivity of the Ge structure array as a function of the free charge-carrier density. The delay time axis is assigned by considering a recombination time of 288 ps. The white lines indicate no reflectivity change whilst the black line follows the Ge-Si antenna resonance. The simulation can be compared with the experimental data range (dashed) by considering a slight blue-shift of the response. The color bar is valid for both graphs.

A more quantitative description of the antennas activation as measured in the experiments, together with a rigorous assignment of the observed optical features to the different antenna resonances, has been derived from electromagnetic simulations performed with a frequency-domain method known as the Rigorous Coupled Wave Analysis (RCWA) [44]. This approach exploits a decomposition of the field in a Fourier basis and the scattering matrix approach to obtain the mode amplitudes in the different layers. TM-polarized reflection spectra have been simulated for all angles between 71° and 81° and then averaged by using a Gaussian function to take into account the focusing conditions of the experiments. The dielectric constant of Ge used for the simulations was calculated considering the effective Drude response of a plasma constituted by both electrons and holes and considering their respective effective masses. In particular, we calculated that the valence band contribution is due to both heavy and light holes. In detail:

$$\upsilon_p = \frac{1}{2\pi}\sqrt{\frac{Ne}{m_L^* \varepsilon_0} + \frac{N_{hh} e}{m_{hh}^* \varepsilon_0} + \frac{N_{lh} e}{m_{lh}^* \varepsilon_0}}$$

where $N$ is the number density of electron-hole pairs, $e$ the fundamental electron charge, $\varepsilon_0$ the free space permittivity and $m^*$ the effective mass of electrons in the L-valley of the conduction band ($m_L^* = 0.12 m_0$) and for heavy and light holes in the valence band ($m_{hh}^* = 0.33 m_0$ and $m_{lh}^* = 0.043 m_0$, respectively). The densities for heavy and light holes sum up to $N$ but their distribution is not equal given the different dispersion of the respective bands. We calculate the ratio between the two numbers to be:

$$\frac{N_{hh}}{N_{lh}} = \left(\frac{m_{hh}^*}{m_{lh}^*}\right)^{\frac{3}{2}}.$$

In addition, we also considered the crucial contribution of low-frequency absorption from the split-off hole band to the emptied states at the Γ point via Lorentzian line shapes for the evaluation of the effective dielectric function. The exact position of these transitions in the energy spectrum depends on the carrier concentration and dynamically red-shifts while electrons and holes recombine. It is worth noting that our modeling approach does not make use of any fitting parameters.

Fig. 4(b) presents the results of the calculated reflectivity as a function of the free carrier density. The horizontal axis is plotted in a logarithmic scale to mimic the recombination dynamics and to allow for a more direct comparison with the experimental trace in the time domain. We notice that the simulation reproduces the strong modulation of the reflectivity around the plasmonic resonance and that describes the shift towards longer wavelengths for lower carrier densities. The calculation qualitatively reflects the experimental observations with only a slight blue shift of the spectral response that might be due to the anharmonicity of the bands and the consequent non-constant effective masses of the carriers. Nevertheless, the assistance of the finite-difference time domain simulations allows us to establish that our experiments track the Ge-Si lower-energy plasmonic response (black line in Fig. 4(b)) at a frequency of approximately 24 THz. The black solid line in Fig. 4(a) follows this evolution and is calculated by tracking the signal amplitude that is expected by the simulation for a given carrier density. Interestingly, at early times we can observe a complex, quasi-stationary evolution of the resonance frequency followed by its monotonic red-shift to below 19 THz within 400 ps. The effect occurring in the first 10 ps is in nice qualitative agreement with the electromagnetic simulations reported in Fig. 4b and might be additionally affected by initial non-equilibrium dynamics in the carrier distribution. The Ge-air resonance lays at higher energies and is not depicted in the figures.

A different perspective for active control of the antenna array on the picosecond time scale exploits the so-called Rayleigh-Woods anomalies [45, 46, 47], i.e. the abrupt discontinuities observed in the transmission and reflection spectra of plasmonic arrays at correspondence of a diffraction orders. Such steep spectral features, that have important practical application e.g. for MIR filters [47, 48], can also be modulated in intensity by the pump pulse, while their spectral position is fixed by the array geometry and therefore does not shift during the transient evolution of the plasma frequency.

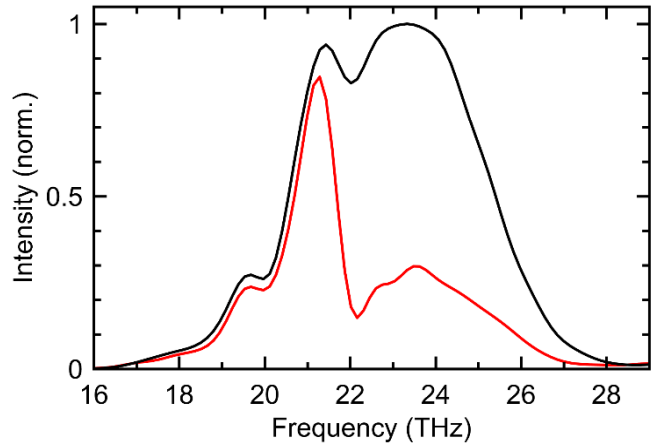

Figure 5 (color online): The reflected intensity spectrum of the 3-µm single-rod antenna array measured with (red line) and without (black line) optical excitation. This sample displays a distinct and steep Rayleigh-Woods anomaly signature at 22 THz that becomes significant after the optical excitation.

In the antenna array constituted by single 3-µm-long rods with 7 µm periodicity we observe (Fig. 5) a sharp variation of the reflectivity that is strongly suppressed above 22 THz upon photoexcitation with the near-infrared pump pulse. Such a system can therefore act as an optically controlled, ultrafast MIR filter whose response can be further optimized and engineered by increasing the complexity of the array geometry [47, 48].

In conclusion, we have demonstrated that it is possible to optically activate intrinsic germanium antennas and establish localized plasmonic resonances that can access the MIR spectral range. This approach gives full control over the field enhancement and confinement of MIR light in the sub-diffraction limit and at the ultrafast timescale. Our experiments lay the foundation for active plasmonic nanosystems that are particularly appealing for the broadband identification of molecules through their vibrations in the fingerprint region. Crucial is the prospect to exploit established all-semiconductor technologies for direct on-chip integration of innovative sensing devices driven by compact picosecond lasers. In addition, the optical switching of near-field optics paves the way for novel approaches in fundamental and semiconductor material sciences. For example, the all-optical manipulation of optoelectronic devices, such as integrated MIR waveguides, detectors and filters will be possible.

**Acknowledgments**

The research leading to these results has received funding from the European Union's Seventh Framework Programme under grant agreement n°613055. We also acknowledge the support of the Deutsche Forschungsgemeinschaft (DFG) through the Emmy Noether Program (BR 5030/1-1).